\documentclass[12pt,preprint]{aastex}

\newcommand{\lya}{Ly$\alpha$}

\newcommand{\tnm}{\tablenotemark}
\newcommand{\solar}{$_{\sun}$}
\newcommand{\teff}{$T_{eff}$}
\newcommand{\logg}{log$\,g$}
\newcommand{\cgen}{\sc{cmfgen}\upshape}
\newcommand{\tlus}{\sc{tlusty}\upshape}
\newcommand{\savcod}{Savage \& Code}
\newcommand{\savpan}{Savage \& Panek}

\shorttitle{\ion{H}{I}~\lya\ EW's of stellar populations}
\shortauthors{Pe\~na-Guerrero \& Leitherer}

\received{}
\begin{document}

\title{\ion{H}{1}\ Lyman-alpha equivalent widths of stellar populations}

\author{Mar\'ia A. Pe\~na-Guerrero\footnotemark[1]}
\email{pena@stsci.edu}

\and

\author{Claus Leitherer\footnotemark[1]}
\email{leitherer@stsci.edu}

\footnotetext[1]{3700 San Martin Drive, Baltimore, MD, 21218, USA}

\begin{abstract}
We have compiled a library of stellar \lya\ equivalent widths in O and B stars using the model atmosphere codes \cgen\ and \tlus, respectively. The equivalent widths range from about 0 to 30 \AA\ in absorption for early-O to mid-B stars. The purpose of this library is the prediction of the underlying stellar \lya\ absorption in stellar populations of star-forming galaxies with nebular \lya\ emission. We implemented the grid of individual equivalent widths into the Starburst99 population synthesis code to generate synthetic \lya\ equivalent widths for representative star-formation histories. A starburst observed after 10 Myr will produce a stellar \lya\ line with an equivalent width of $\sim$ -10$\pm$4 \AA\ in absorption for a Salpeter initial mass function. The lower value (deeper absorption) results for an instantaneous burst, and the higher value (shallower line) for continuous star formation. Depending on the escape fraction of nebular \lya\ photons, the effect of stellar \lya\ on the total profile ranges from negligible to dominant. If the nebular escape fraction is 10\%, the stellar absorption and nebular emission equivalent widths become comparable for continuous star formation at ages of 10 to 20 Myr.
\end {abstract}


\section{Introduction}\label{intro}
Galaxies with active star formation host large populations of massive, hot, young O and B stars which ionize the surrounding interstellar medium (ISM). The resulting nebular recombination spectrum includes numerous strong emission lines which are widely used as star-formation tracers \citep{ken12}. Among these lines, Lyman-alpha (\lya) deserves special attention. \lya\ photons are expected to be produced when the opacity approaches infinity for those photons associated with the base level (Case B). Hence, if Case B applies, strong \lya\ emission is expected, and this line can be detected even at the highest redshifts \citep[i.e.][and references therein]{par67, mal02, ste05, bro11, san13}. 

At high redshift \lya\ is often the only detectable emission line in the optical and near-infrared, and is therefore widely used as an indicator of star formation \citep[i.e.][]{mal02, sca09, dij12}. If the continuum light is detected, the \lya\ equivalent width (EW) can give valuable clues on population ages and the stellar initial mass function (IMF) \citep{lei10}. If Case B applies, the theoretically predicted values of EW are expected to vary from 50 to $\sim$240 \AA\ in star-forming galaxies with ages of less than ~100 Myr \citep[i.e.][and references therein]{cha93, lau13}. However, \lya\ turns out to be a complex star-formation tracer. In most cases the observed line strength is less than expected from Case B recombination because the effects of radiation transfer have not been taken into account \citep{han06}. The assumption of \lya\ being a pure recombination line in a gaseous medium is too simple. \citet{mei81}, \citet{har88}, \citet{neu90}, and \citet{cha93} considered the effects of dust on the \lya\ radiative transfer. Resonant scattering by atomic hydrogen significantly increases the path length for \lya\ photons and consequently the likelihood of \lya\ destruction by dust, leading to much lower line strengths. On cosmological scales, models incorporating a redshift-dependent intergalactic dust obscuration \citep{hai99, mal04} predict \lya\ space densities that are broadly consistent with ``semi-analytical'' models connecting \lya\ emitting galaxies and dark matter halos. 

Although there is consensus that \lya\ is regulated by dust, no clear correlation has been found between dust content and \lya\ EW at low redshift (z) \citep[]{gia96, han06}. However, at $z\sim3$ the picture seems to be rather different. \citet{sha03} found a significant correlation between UV continuum extinction and \lya\ EW. They suggest that this difference between the low and high redshift could be due to either differences in the geometry of dust in the neutral gas, or small sample statistics biases at low redshift. Furthermore, \citet{lau13} found that different ISM morphologies are unlikely to account for different \lya\ escape probabilities. Instead, they propose an increase of \lya\ radiation due to cold accretion and/or anisotropic escape. Most importantly, the kinematic properties of the ISM may very well be the dominant escape or trapping mechanism for \lya\ \citep{kun03, fuj03, mas03, dij10}.

In addition to physical processes related to the ISM, the observed \lya\ emission can be modified by corrections for any underlying stellar \lya. Depending on the stellar population properties, these corrections can be quite substantial and even cancel out the nebular emission. \citet{val93} suggested stellar \lya\ absorption from an aged population as the reason for the dearth of \lya\ emitters among local starburst galaxies. 

The stellar \lya\ cannot be fully constrained observationally since even the closest O stars in our Galaxy have ISM \ion{H}{1} column densities high enough to produce strong interstellar \lya, which masks their total intrinsic stellar line. In this work we retrieved publicly available theoretical spectra of O and B stars generated with \cgen\ \citep{hil98, hil11} and \tlus\ \citep{hub95, lan03, lan07}, respectively, and determined 276 individual \lya\ EW's. The EW's are used as a library for implementation in the Starburst99 synthesis code \citep{lei99, lei09}. The goal of our work is to predict the behavior of the stellar \lya\ as a function of population properties. While understanding the behavior of stellar \lya\ is crucial in its own right, we are mainly motivated by the need to evaluate the effect of the stellar contribution to the total \lya\ emission in star-forming galaxies. This work is organized as follows. In Section \ref{method}\ we present a description of the models used for the new theoretical library, and we describe how we determined EW's. In Section \ref{comparison} we compare our determinations with observations of individual O and B stars, and in Section \ref{popsyn} we describe the implementation of the new library in the Starburst99 code and the resulting EW's of typical populations. In Section \ref{conc} we present our conclusions.

\section{Methodology}\label{method}
Our goal is to predict the stellar \lya\ line strength of a population of young stars whose ionizing photons give rise to the observed nebular \lya\ emission. The relevant stars are of spectral types O and B. Later types contribute negligible light at 1216 \AA\ in a population with OB stars present. Alternatively, a single, evolved population at an age older than $\sim100$ Myr (when A stars start to dominate) would not be observed as a \lya\ emitter because of the dearth of ionizing photons.

Strong stellar winds are a defining characteristic of OB stars. The spectral signatures of these winds are blueshifted absorption, broad emission, or P Cygni-type profiles \citep{pul08}. \lya\ in particular, is susceptible to wind effects due to its formation depth far above the photosphere. This was already recognized during the early efforts of non-Local Thermal Equilibrium (non-LTE) modeling by \citet{kle78} who emphasized that the strength of \lya\ in luminous O stars is never significant due to canceling effects of the emission and absorption components.

In this work we utilize stellar spectra of O stars computed with the \cgen\ atmosphere code, which was designed to solve the radiative transfer and statistical equilibrium equations in hot stars in spherical geometry \citep{hil01}. The models account for non-LTE and are fully blanketed. The \cgen\ code was originally written for modeling the strong winds of Wolf-Rayet stars \citep{hil87}, but has since been generalized to be applicable to O stars as well. Pre-calculated grids of O-star models are readily available in various databases so that there is no need to generate additional models. We surveyed the available spectra of O stars from the Pollux database \citep{pal10}, and retrieved 46 \cgen\ O star spectra of solar metallicity, microturbulent velocity of 5 and 10 km/s, effective temperatures (\teff) ranging from 27,500 to 48,530 K, and surface gravity (\logg) ranging from 3.0 to 4.25. The mass-loss rate used in this O-grid ranged from 4.85-8.34$\times10^{-6}$ M\solar/yr, and has been adopted to match the observed rates. A $\beta$-velocity law \citep{lam99} was adopted, with $\beta=0.8$. The O-grid available on the \cgen\ webpage\footnote{http://kookaburra.phyast.pitt.edu/hillier/web/CMFGEN.htm}\ is a subset of that contained in the Pollux database and therefore provides no additional new data (the O-grid in the \cgen\ website contains only 23 models of O stars). 

Hot-star winds scale with luminosity and (to a smaller degree) with \teff\ \citep{vin07}. As a result, wind effects become less pronounced in B stars when compared to O stars. B star spectra show less evidence for winds, except for the most luminous supergiants, which are rare by number in a typical stellar population. Given this situation, we opted for B-star models based on the \tlus\ atmosphere code \citep{lan07}. Like \cgen, \tlus\ is a fully metal-blanketed non-LTE code. In contrast to \cgen, \tlus\ adopts a static, plane-parallel geometry, which does not account for stellar-wind effects above the photosphere. \tlus\ is an excellent representation of B main-sequence stars and of evolved B stars with moderate winds. Modeled spectra of B stars were obtained from the \tlus\ website\footnote{http://nova.astro.umd.edu/index.html}. We used the BSTAR2006 grid of B stars with solar metallicity. We retrieved 212 modeled B star spectra with microturbulent velocities of 2 and 10 km/s, \teff\ ranging from 15,000 to 30,000 K, and \logg\ ranging from 1.75 to 4.75. The specific characteristics of the models are presented in Lanz \& Hubeny.

Since winds are not completely absent in B-type stars, we need to be concerned about the applicability of the static, \tlus\ models. \citet{hea06} provide an extensive discussion of the trades between the \cgen\ and \tlus\ codes. While the preference for spherical, expanding atmospheres in O stars is obvious, it is preferable to use sophisticated photospheric models such as \tlus\ for stars with moderate or weak winds rather than rely on spherical models with a simplified physics of the deep, quasi-static layers. This justifies our choice of \cgen\ for O stars, and \tlus\ for B stars.

\citet{mas13} found that the choice of different microturbulent velocities has little effect on heavy element lines and no effect on the hydrogen lines. We measured the difference in the determined EW's from models with different microturbulent velocities. We found that the mode on this difference was $\sim$3 \AA, and the minimum and maximum differences were 0.1 and 7.2 \AA, respectively. We used the microturbulence velocity of 10 km/s unless this value was not available. In Table \ref{tmods} we give a summary of the models compiled for this work. There is a total of 258 model spectra considering all microturbulent velocities. The parameters cover the full observed range of \teff\ and \logg\ in the upper Hertzsprung-Russell diagram.

In order to obtain determinations of \lya\ EW's\footnote{The sign convention used in this work is positive for emission lines and negative for absorption lines.} comparable to those obtained from observations, we rebinned the spectra from the extremely high resolution of the models ($\Delta\lambda$ of 0.060 \AA\ for \cgen\ and 0.005 \AA\ for \tlus\ at \lya) to $\Delta\lambda$ of 0.465 \AA. This was done using a cubic spline function. This value is an average of $\Delta\lambda$ of 0.75 and 0.18 \AA, the typical resolutions of the two spectrographs on board of HST, the Space Telescope Imaging Spectrograph and Cosmic Origins Spectrograph, used for observing extragalactic objects. 

Our determinations of EW's were done with a simple flux over continuum integration code written in Python. For the \tlus\ modeled atmospheres, both the continuum and the lines spectra were retrieved from their webpage. Hence, the EW determination was straight forward since the continuum location is predefined. However, for the Pollux retrieved SEDs, only the lines spectra were available. As expected, the EW determination changed with the height where the continuum was placed. The continuum was set to be a straight line between two points free of lines in a wavelength range from 1100 to 1300 \AA. To test the validity of this method we used the small grid of O stars from the \cgen\ website that do give both the continuum and the lines files. We retrieved these files and normalized the spectra (divided the lines by the continua files). We then determined the EW's with the Python code integrating through 10 \AA\ centered at \lya. For the purpose of comparison we will call these determinations the benchmark EW's. We then used only the lines files to fit our straight line continuum method and determined the EW's with the same Python code. The resulting EW's have on average an error of 1\%\ and up to 5\%\ with respect to the benchmark EW's.

\begin{figure}
\begin{center}
\includegraphics[angle=0,scale=0.7]{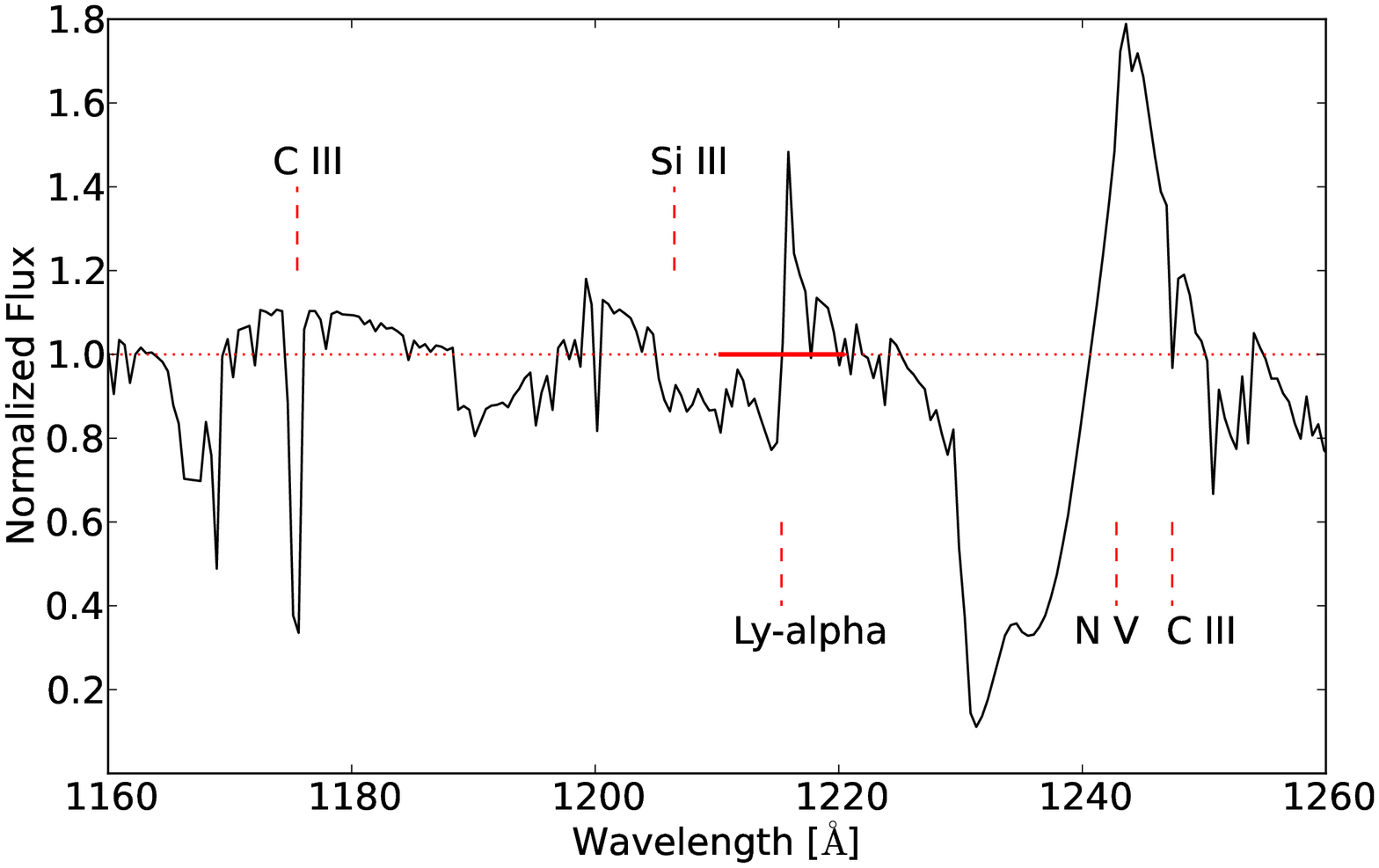}\\
\caption[fig_Ostar.eps]{
Theoretical spectrum of a \cgen\ O star with effective temperature of 40,000 K, \logg\ of 3.5, and micro turbulent velocity of 10 km/s. The red dashed vertical lines mark the rest wavelength of the most prominent lines. The red horizontal dotted and solid lines mark the continuum and integration window used for the ``simple integration" method (see Section \ref{method}), respectively.
\label{fOstar}}
\end{center}
\end{figure}

\begin{figure}
\begin{center}
\includegraphics[angle=0,scale=0.7]{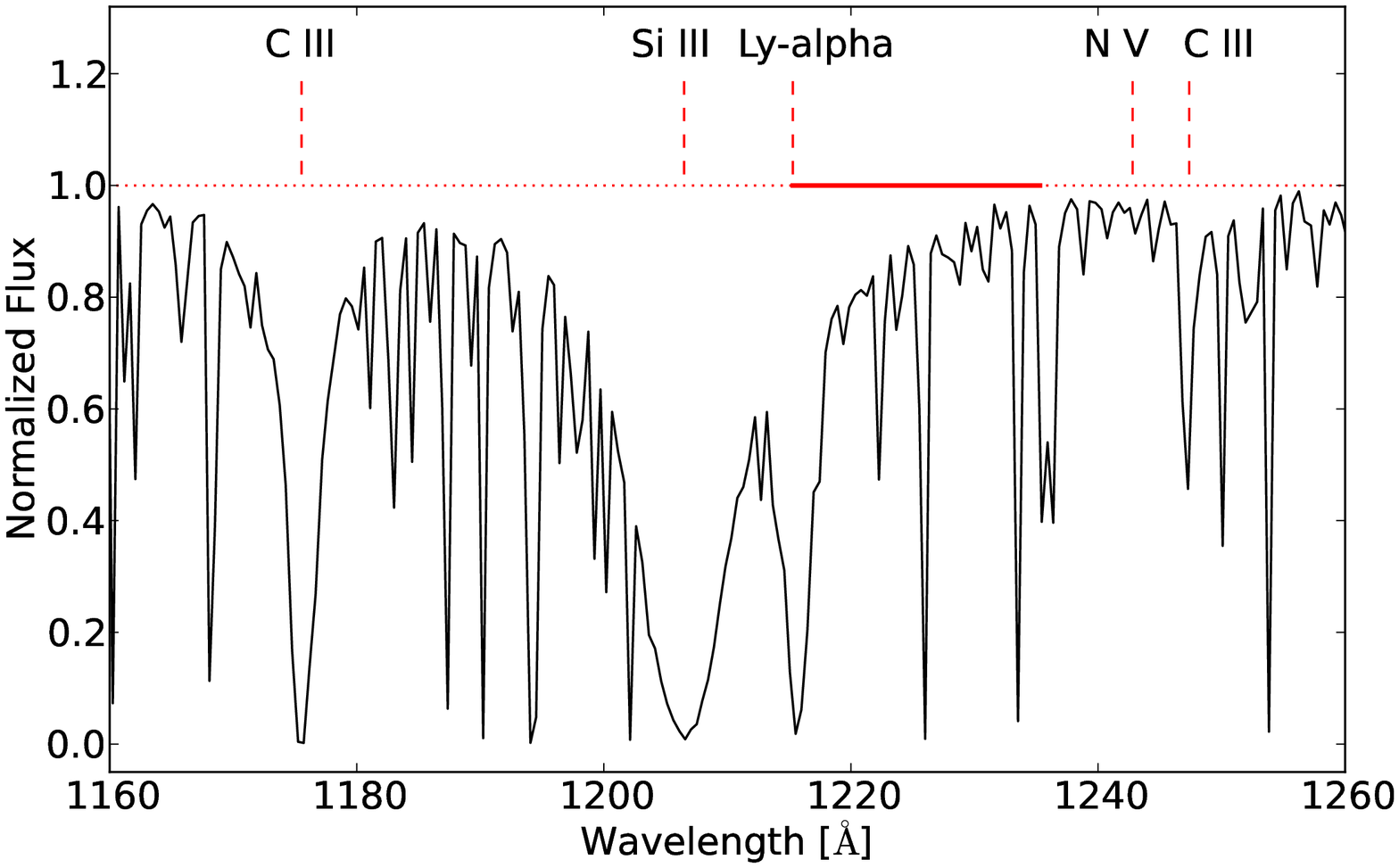}\\
\caption[fig_Bstar.eps]{
Theoretical spectrum of a \tlus\ B star with effective temperature of 18,000 K, \logg\ of 4.25, and micro turbulent velocity of 2 km/s. The red dashed vertical lines mark the rest wavelength of the most prominent lines. The red horizontal dotted and solid lines mark the continuum and integration window used for the ``half integration" method (see Section \ref{method}), respectively.
\label{fBstar}}
\end{center}
\end{figure}

Figures \ref{fOstar} and \ref{fBstar} show representative spectra of O and B stars, respectively. In both figures the most prominent lines are marked for reference, and the red horizontal line indicates the position of the continuum and the EW integration range for each star type. These reference lines are: \ion{C}{3} 1175 \AA, \ion{Si}{3} 1206 \AA, \ion{H}{1} \lya\ 1216 \AA, \ion{N}{5} 1240 \AA, and \ion{C}{3} 1247 \AA. For O stars the spectra of most models show clear P Cygni profiles, indicating the combination of photospheric absorption and high density stellar winds due to radiation pressure \citep{lei95}. In the spectra of luminous O stars and even in the earliest B stars, the \ion{N}{5} line can be quite prominent (and in some cases blend with \lya), though its strength decreases for later B stars \citep{wal85}. In turn, the \ion{Si}{3} line does not play a major role in the spectra of O stars but in B stars this line can easily blend with \lya, since the \lya\ width increases considerably for later B type stars \citep{wal95}. Since in this study we aim to determine the stellar contribution of the observed \lya\ EW, we want to avoid most of the possible blendings with \lya. We therefore chose different EW integration ranges for O and B stars.

For the modeled O stars, the integration was performed over 10 \AA\ centered on the rest frame wavelength of the \lya\ line (strictly 1215.67 \AA). The EW's of these stars were on average in absorption and mostly close to 0 \AA\ due to the strong P Cygni wind profiles (see Figure \ref{fOstar}). Hence, the chosen wavelength range of 10 \AA\ allowed us to account for the full width of \lya, including wings and the P Cygni profiles, if present. 

For the modeled B stars, the integration was done over 40 \AA\ centered on the rest frame wavelength of \lya\ in order to account for the broader \lya\ absorption. For clarity we will call this determination ``simple integration'' EW. Nonetheless, the \ion{Si}{3} resonance line at 1206.5 \AA\ is an important source of line blending for B stars \citep{sav74}. In order to correct for the contribution of this line to \lya\ we assumed that \lya\ was symmetric, and avoided the part of the spectrum where \ion{Si}{3} and \lya\ overlap. We integrated the EW from the rest wavelength of \lya\ to \lya+20 \AA\ and then multiplied this EW by 2. We call this determination ``half integration" EW. 

\citet{sav74} found that the typical contribution of the \ion{Si}{3} line to \lya\ is about 4 \AA. When comparing the \lya\ EW determinations through ``simple integration'' EW to the ``half integration'' EW, we found that the difference between the first and the second EW determination is on average 3 \AA\ but peaks between \logg\ of 2.0 and 2.50 to about 7 \AA\ below an effective temperature of 21,000 K. We chose the ``half integration" EW to better represent the \lya\ EW since the \ion{Si}{3} line is a clear systematic feature in the modeled B star spectra (see Figure \ref{fBstar}). The resulting EW determinations for B stars are all in absorption and range from 1 to 32.5 \AA.

\begin{figure}
\begin{center}
\includegraphics[angle=0,scale=0.7]{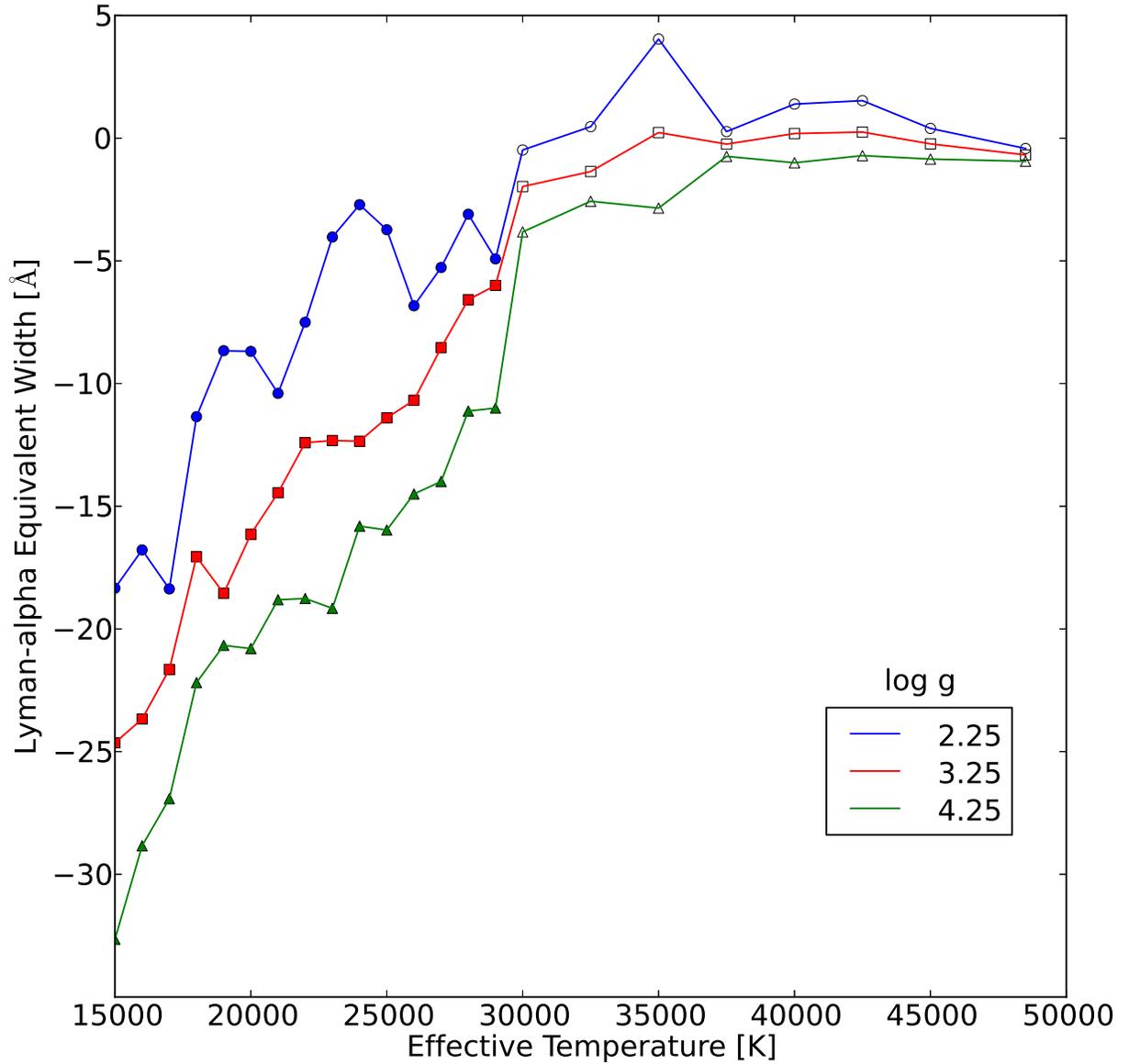}\\
\caption[fig_EWvsTeff.eps]{
Plot of \lya\ EW versus effective temperature. Three representative curves are presented. The top curve (blue circles) represents the \lya\ EW for stars with a \logg\ of 2.25, the middle curve (red squares) for stars with a \logg\ of 3.25, and the lower curve (green triangles) for stars with \logg\ of 4.25. Filled symbols represent \tlus\ stars, while open symbols represent \cgen\ stars.
\label{fEWvsTeff}}
\end{center}
\end{figure}

For \teff\ of 15,000 to 30,000 K we used the \tlus\ models and for \teff\ $>30,000$ K we used the \cgen\ models. We did a linear interpolation to obtain the EW's for those values of \teff\ and \logg\ that were not included in the model spectra we retrieved. The final EW's adopted in this work are presented in Table \ref{tewtg}. Figure \ref{fEWvsTeff} shows the trend of our EW determinations with respect to \teff\ for different values of \logg. We attribute the slight jump in this figure to the different treatments of the stellar winds used in \tlus\ and \cgen\ codes. Nonetheless, in order to check for consistency between both codes, we studied in the differences between EW measurements in the overlapping region (i.e. \teff\ between 27,000 and 30,000 K). We found that the mode and average difference of the \cgen\ and the \tlus\ EWs is about 7 \AA, with a maximum difference of 9 \AA\ at \teff\ of 30,000 K and \logg\ of 3.75.

The behavior shown in Figure \ref{fEWvsTeff} is consistent with what is expected from observations. For both O and B stars, the intensity and width of the ultraviolet features is closely related to the luminosity type \citep{wal95}. Spectra of O stars show a P Cygni profile in some lines (late-B stars for metal lines) when the absorbing ions column density is larger than about $10^{15}$ ions per cm$^{-2}$ \citep{lam99}. This is precisely the overlapping region in our EW measurements, from 27,000 to 30,000 K. The P Cygni profiles for \lya\ become progressively smaller as the \teff\ decreases, hence making the EW progressively differ from zero (i.e. a greater absorption as the wind emission becomes smaller). As \teff\ keeps decreasing, the \lya\ absorption becomes wider.

In order to explore how non-solar chemical compositions affect the \lya\ EW, we obtained spectra of \tlus\ of B stars with \teff\ of 30,000 K and \logg\ of 4.25, and with 4 different metallicities besides solar. These stars are also described in \citet{lan07}. We determined the EW's with the ``half integration" method. The determined \lya\ EW's for metallicities (in units of Z/Z\solar) of 2, 1, 1/2, 1/5, and 1/10 were: -15.38, -12.87, -10.43,  -9.85, and -6.30 \AA, respectively. The overall shape of the spectrum changes very little with metallicity, i.e. the lines around \lya\ become less prominent, making it easier to detect \lya\ even considering that the EW is also getting smaller. This behavior is as expected since: (i) the EW is defined as the wavelength range over which the continuum around the line must be integrated to produce the same intensity as the observed line \citep{lal11}, and (ii) the effect of metallicty on other hydrogen lines such as the Balmer lines is also rather weak if the \teff\ is higher than 7,000 K \citep{gon99}.

\section{Comparison with Observations of O and B stars}\label{comparison}
In order to gain confidence into our set of \lya\ EW's, we surveyed the literature for available observations of stellar \lya. There is a rich body of data dating back to the early observations by OAO-2 \citep{sav72}, OAO-3 \citep{boh78}, and IUE \citep{dip94a, dip94b}. We compared our theoretical \lya\ EW determinations with those from observations of O and B stars presented in \citet{sav70} and \citet{sav74}. The observations presented in \savcod\ and \savpan\ were taken with OAO-2, whose spectrograph had a resolution of 15 \AA. The relatively low resolution implies that there could be contamination with \ion{Si}{3} on the short-wavelength side of \lya\ and possibly with \ion{N}{5} on the long-wavelength side, depending on the extent of the wings of \lya. 

We adopted the EW's as published by \citet{sav70} and \citet{sav74}, with the sign convention used through out this work. Contamination of stellar \lya\ by the interstellar \ion{H}{1} absorption is a serious concern, in particular for O stars whose larger average distances (when compared to B stars) lead to interstellar neutral hydrogen column densities $N_H$ in excess of $10^{20}$ cm$^{-2}$. As a result, the measured \lya\ EW's in essentially all O stars are not stellar but interstellar. Since the derived lower limits of the stellar contribution to the total \lya\ would provide little insight into the validity of our models, we did not consider stars with \teff $> 35,000$ K for our comparison. This cut-off corresponds to spectral type O9. In Table \ref{tewobs} we present the selected stars together with the pertinent parameters. Columns 1 and 2 list the designations. The spectral types are in column 3. \teff\ (column 4) was assigned using the calibration of \citet{con08}. Observed EW values are in column 5.   

\begin{figure}
\begin{center}
\includegraphics[angle=0,scale=0.66]{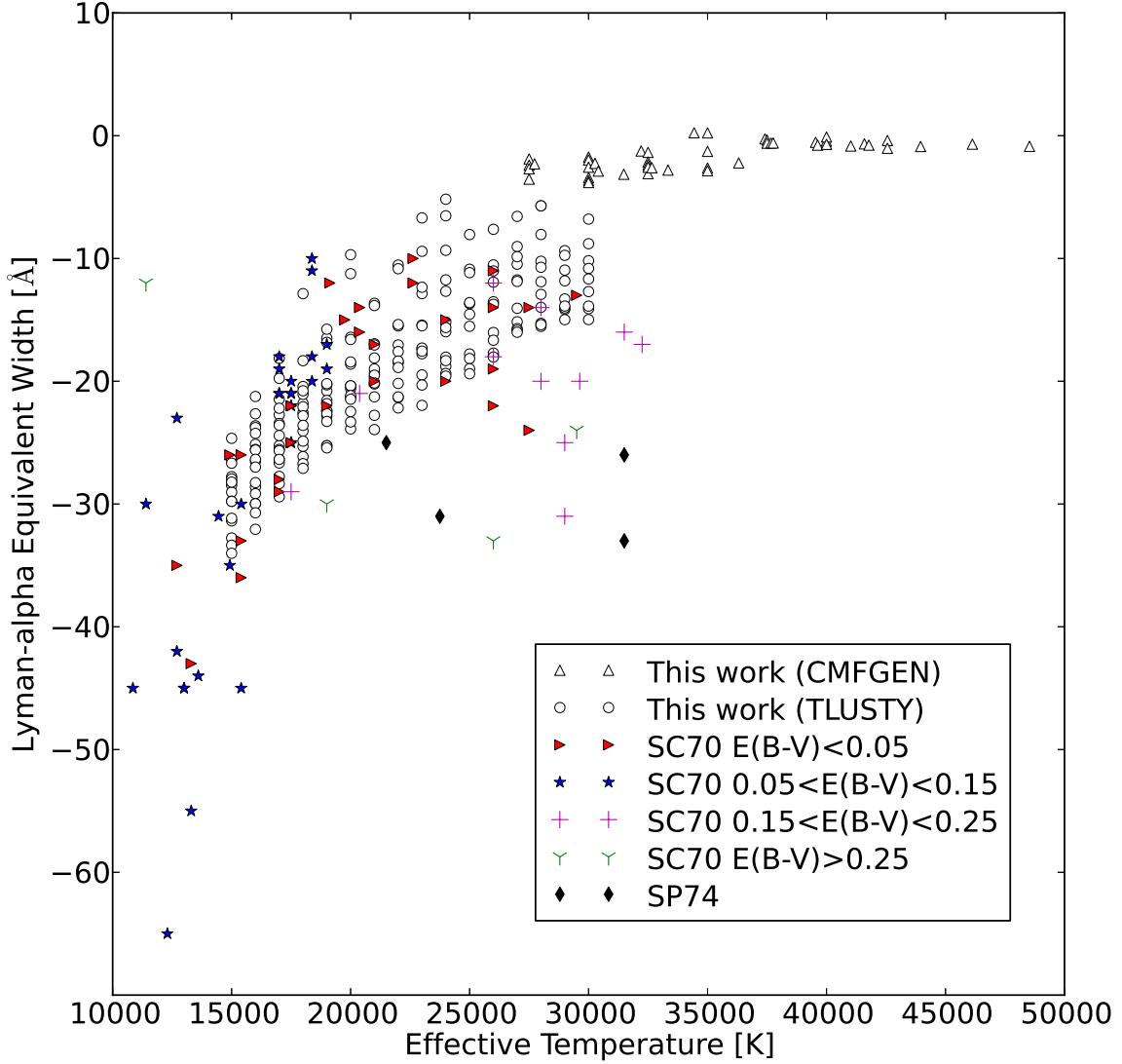}\\
\caption[fig_CompVsObs.eps]{
Comparison between our \lya\ EW determinations and those from observations, as a function of effective temperature. Symbols represent the following: open up triangles are EW determinations of this work, from \cgen\ models, open circles are EW determinations from \tlus\ models, red rotated triangles, magenta crosses, black diamonds, and green three-pointed stars are EW obtained from observations from \citet{sav70}, SC70, for different values of $E(B-V)$, and blue stars are EW obtained from observations from \citet{sav74}, SP74. SC70 measurements are blended with the interstellar \lya\ line and with neighboring lines, and SP74 are corrected for both ISM and blending with the \ion{Si}{3}.
\label{fCompVsObs}}
\end{center}
\end{figure}

In Figure \ref{fCompVsObs} we show the comparison between the observed and theoretical EW values as a function of stellar effective temperature. The models display a near-monotonic trend with \teff, asymptotically reaching about zero \AA\ at \teff\ $> 35,000$ K. At a given \teff, a larger \logg\ results in stronger \lya\ absorption. The observed values track the models very well at the lowest temperatures but tend to be more negative (stronger absorption) at higher temperatures (corresponding to early B stars). We interpret this as not due to a model failure but rather being caused by residual interstellar contamination. In support of this suggestion we indicated the measured $E(B-V)$ values for each observation, taken from \citet{sav70} and \citet{sav74}. There is a clear trend of the largest discrepancies between models and data occurring for the highest $E(B-V)$ values. Since there is a well-established relation between $E(B-V)$ and interstellar $E(B-V)$ \citep{dip94a}, we conclude that the majority of the observational points below the models are in fact  just lower limits to the stellar contribution.

The outcome of the comparison between the modeled and observed EW's is quite reassuring. While there are few observational constraints on the \lya\ EW in O stars, the intrinsic \lya\ in the hottest stars is expected to be weak because of the counteracting effects of photospheric absorption and wind emission. Therefore these stars would make only minor contribution to the total stellar absorption in a representative population. B stars, on the other hand, when present in a population can contribute substantially, and their modeled \lya\ is in very good agreement with the data.

\section{Population Synthesis Models}\label{popsyn}
The \lya\ EW's listed in Table \ref{tewtg} were implemented in the synthesis code Starburst99 \citep{lei99, lei09}. We performed a 2-dimensional interpolation in \teff\ and \logg\ to obtain the EW values for each point in the Hertzsprung-Russell diagram as prescribed by the stellar evolutionary tracks. The input table is essentially complete at the highest temperatures, and no significant extrapolations were required. For \teff\ between 15,000 K and 10,000 K (late B stars) we used eq. (1) of \citet{val93} to approximate the \lya\ EW. No attempts was made to account for EW at even lower \teff, and we simply assumed a featureless continuum for \teff\ $< 10,000$ K. This means our models are no longer valid when A stars begin to contribute to the UV continuum. We do not distinguish between O- and Wolf-Rayet stars when assigning \lya\ EW's to stars in the Hertzsprung-Russell diagram. The underlying assumption is that \lya\ (or the \ion{He}{2} line at approximately the same wavelength) in Wolf-Rayet stars behaves in a similar fashion as in hot O stars.

The nebular \lya\ was calculated assuming standard Case B recombination for an electron temperature of $10^4$ K. For reference, the \lya/H$\alpha$ ratio is 8.7 under these assumptions (Case A would be 11.4).

We created a standard set of synthetic \lya\ EW for evolving stellar populations using the Geneva evolution models with high mass loss at solar chemical composition \citep{mey94}. As a consistency check, we compared our predictions with those published in the literature. A standard Salpeter initial mass function (IMF) has been adopted for these comparisons. \citet{cha93} and \citet{val93} were the first to draw attention to the importance of underlying stellar \lya\ for the interpretation of nebular \lya\ in star-forming galaxies. They both used the original \citet{bru93} evolutionary synthesis models with a correction of stellar \lya\ to predict the net nebular + stellar \lya. The stars evolve along the tracks computed by \citet{mae89} in these models. In Table \ref{teqws} we compare their and our predictions for selected ages and for the two limiting cases of an instantaneous burst and continuous star formation. Also included in Table \ref{teqws} are the predictions of \citet{sch08} who used a modified version of the \citet{sch03} evolutionary synthesis models for predictions of nebular + stellar \lya. The different sets of models are in reasonable agreement. The differences reflect the use of different evolutionary tracks and different model atmospheres. Note in particular that the larger EW at age $5\times10^6$ yr in the Charlot \& Fall and Valls-Gabaud models is the result of different life times in the stellar evolution models. At this age, the nebular \lya\ displays a sharp drop with age, and a slight shift in age causes a vastly different ionizing photon output. 

\begin{figure}
\begin{center}
\includegraphics[angle=0,scale=0.62]{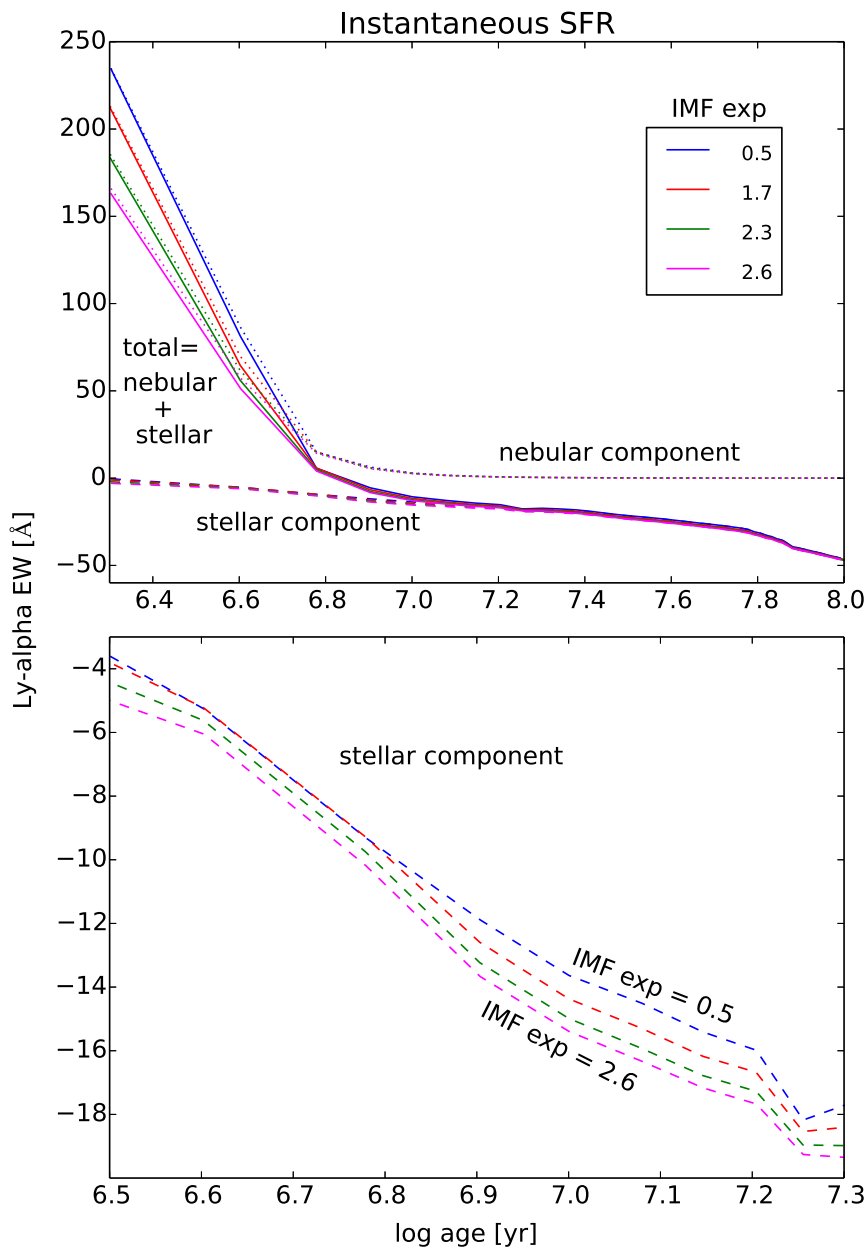}\\
\caption[fig_instSFR.eps]{
\lya\ EW's versus age for an instantaneous burst. This figure shows a series of Starburst99 runs for different IMF exponents, labeled by color from 0.5 (blue) through 2.6(magenta). Salpeter is represented by the green curve. Dashed lines represent the stellar component of \lya\ EW, dotted lines represent the nebular component, and solid lines the total component. The sign convention used for this work is positive for emission and negative for absorption. The upper panel shows the overall behavior of \lya\ up to 100 million years. The bottom panel shows a zoom-in of the stellar component from about 3 to 20 million years.
\label{fInst}}
\end{center}
\end{figure}

\begin{figure}
\begin{center}
\includegraphics[angle=0,scale=0.62]{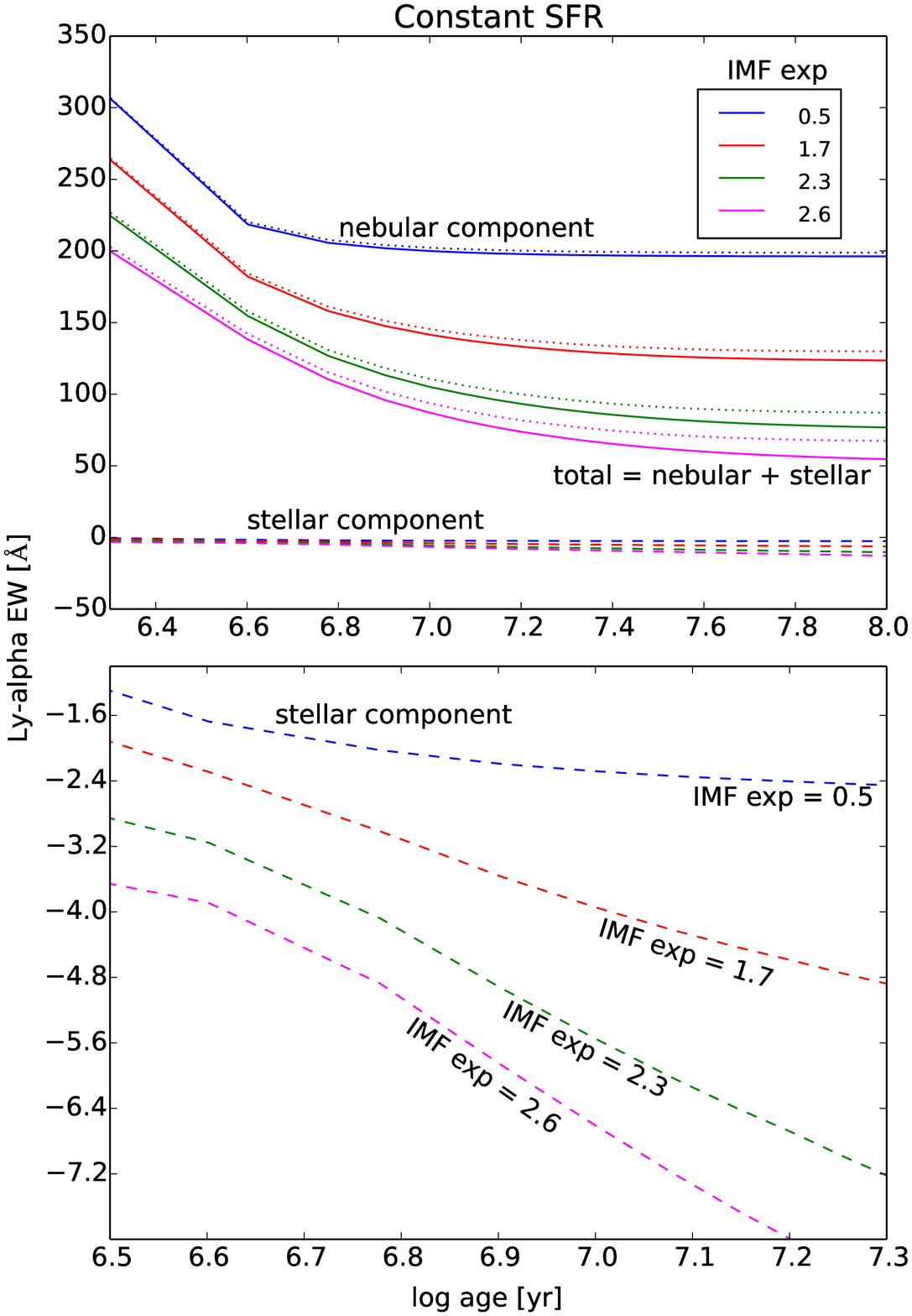}\\
\caption[fig_constSFR.eps]{
Same as Figure \ref{fInst} but for constant star formation rate. 
\label{fConst}}
\end{center}
\end{figure}

We performed a parameter study in order to analyze the impact of different IMFÕs, different ages, and different star-formation histories in the \lya\ EWÕs. The two bracketing cases of an instantaneous burst and constant star formation were considered. Starburst99 assumes a power-law IMF of the classical form $\phi(m) \propto m^{-\alpha}$. In this notation, the classical Salpeter IMF has an exponent of $\alpha = 2.35$. In Figures \ref{fInst} and \ref{fConst} we show the resulting \lya\ EWÕs for different IMF slopes in the case of an instantaneous burst and continuous star formation, respectively.  

We first address the relative importance of the underlying stellar \lya\ at different ages in each of the two cases. The burst models (Fig. \ref{fInst}) fall into two regimes: at young ages (t $\lesssim$ 6 Myr), the nebular part always dominates since copious O stars photons are available and O stars have intrinsically weak \lya. At older ages (t $\gtrsim$10 Myr), the starburst is B-star dominated with few ionizing photons but at the same time intrinsically strong stellar \lya. In the case of continuous star formation (Figure \ref{fConst}), the nebular component always dominates, and the underlying stellar \lya\ is always negligible. As a reminder, this assumes pure Case A recombination for the nebular \lya\, with a 100\% photon escape.

Different IMF exponents, ranging from an extreme, flat IMF of $\alpha=0.5$ to another extreme, steep value of $\alpha=2.6$, only have a minor influence in the {\em stellar} \lya\ EWÕs. The equivalent width varies by about 2 \AA\ for the considered IMF exponents between 0.5 and 2.6 for an instantaneous burst. This is expected for a single population where the stellar flux at a particular wavelength comes from stars over a very narrow mass range. In the continuous case, the IMF sensitivity of the stellar \lya\ is somewhat more pronounced but still quite minor.

Next, we turn to the IMF dependence of the {\em nebular} component of \lya. The IMF has a much more pronounced effect in this case, which can easily be understood from the fact the EW(\lya) is determined by the ratio of early- to late-O stars, which contribute the ionizing photons and the continuum. Varying the IMF slope will change the relative contributions of these species. Therefore the IMF behavior in Figs. \ref{fInst} and \ref{fConst} is not surprising and mirrors that of, e.g., the H$\alpha$ or H$\beta$ equivalent width, which has been extensively discussed in the literature \citep{lei99}. 

\begin{figure}
\begin{center}
\includegraphics[angle=0,scale=0.62]{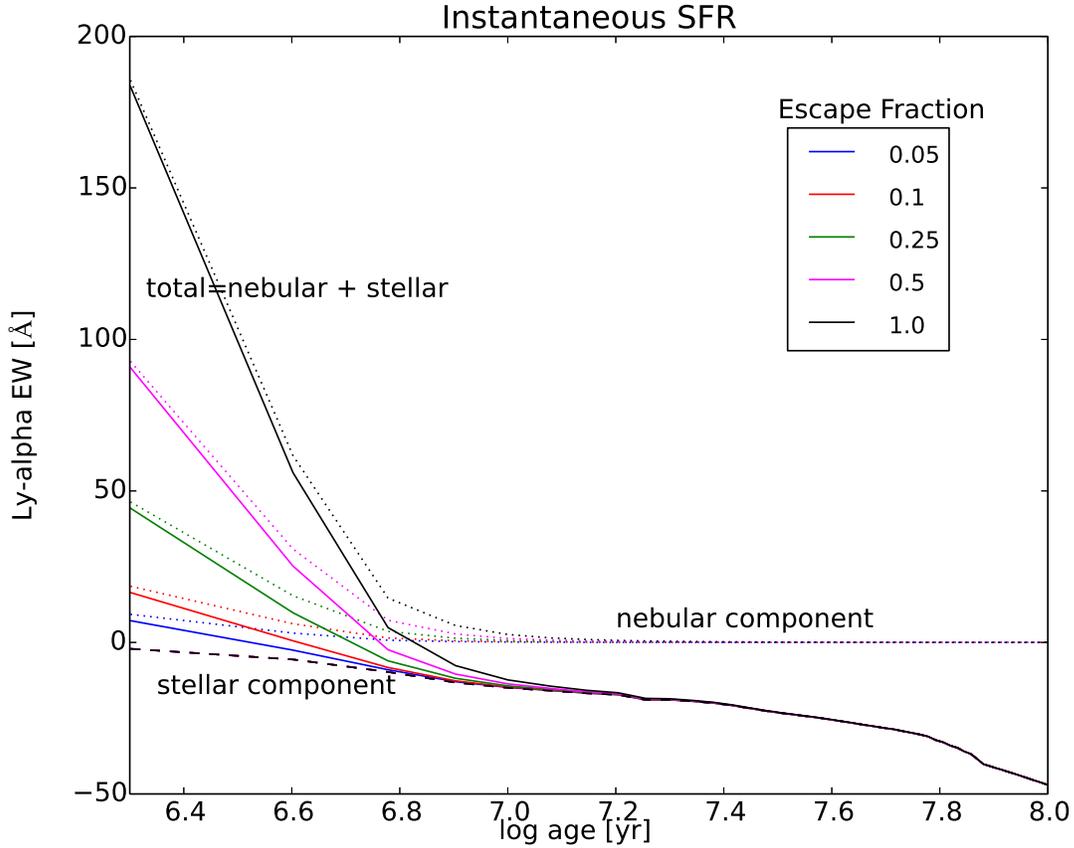}\\
\caption[fig_instSFRdiffeEscErac.eps]{
\lya\ EW's versus age for an instantaneous burst for different \lya\ escape fractions. This figure shows a series of Starburst99 runs for an IMF exponent of 2.3. Dashed lines represent the stellar component of \lya\ EW, dotted lines represent the nebular component, and solid lines the total component. The sign convention used for this work is positive for emission and negative for absorption. \label{fescapeInst}}
\end{center}
\end{figure}

\begin{figure}
\begin{center}
\includegraphics[angle=0,scale=0.62]{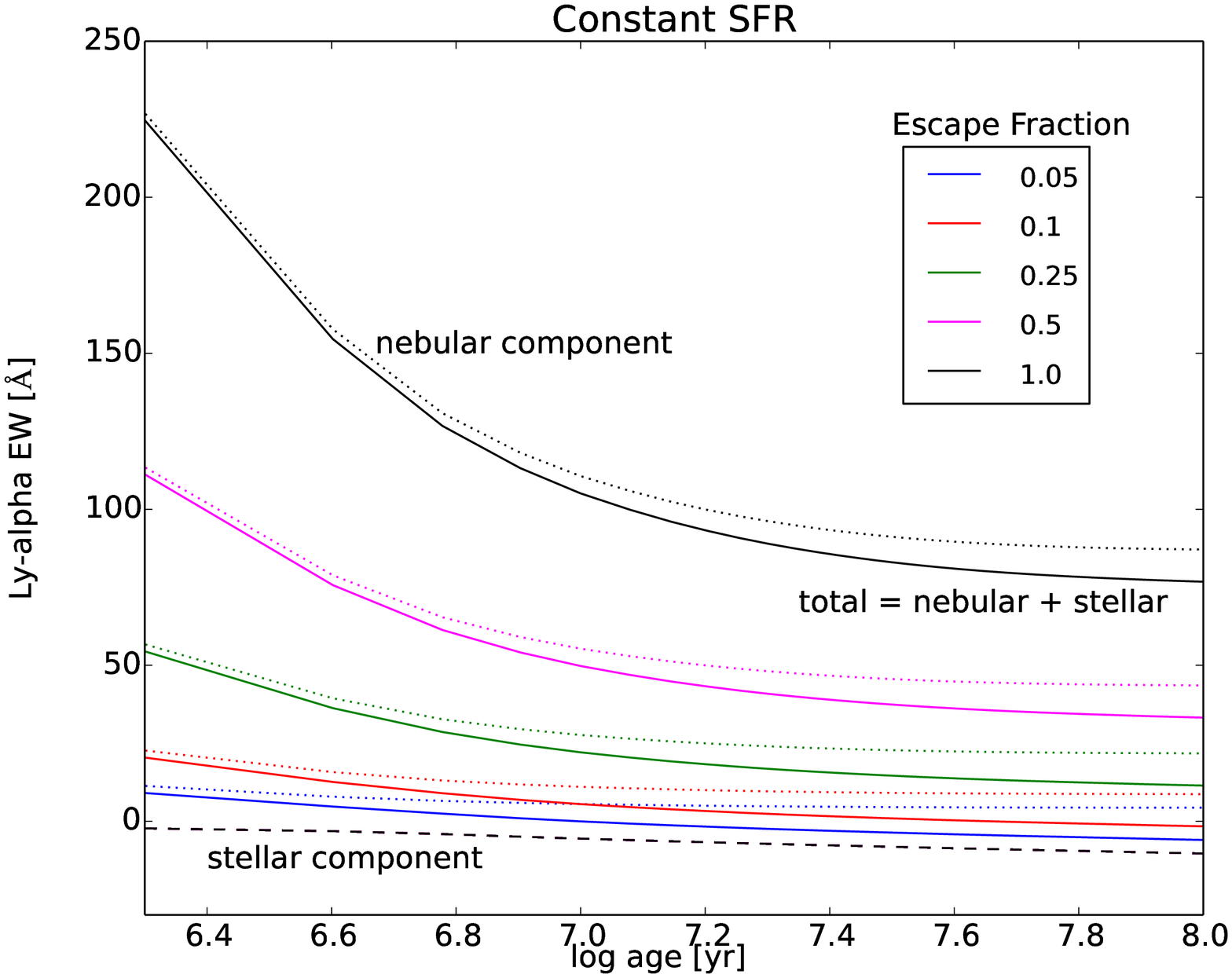}\\
\caption[fConstDiffEscFrac.eps]{
Same as Figure \ref{fescapeInst} but for constant star formation rate. 
\label{fescapeConst}}
\end{center}
\end{figure}

The previous discussion is based on the assumption of a 100\% escape probability of the nebular \lya\ photons. Observations suggest a much lower escape fraction. For instance the \lya\ EWÕs observed in Lyman-break galaxies at redshift $\sim3$ are of order 10 \AA\ \citep{sha03}. Comparison with our continuous models uncorrected for stellar absorption, implies an escape of \lya\ photons of $\sim10$\%. In order to provide a more realistic model grid for comparison with data, we generated models with a Salpeter IMF and nebular \lya\ escape fractions of 5, 10, 25, 50 and 100\%. These models are plotted in Figures \ref{fescapeInst} and \ref{fescapeConst} for instantaneous bursts and continuous star formation, respectively. Because of the reduced nebular emission for escape fractions less than 100\%, the underlying stellar absorption drastically increases in its contribution to the net EW. For instance, if the escape fraction is 10\% the stellar absorption and nebular emission EW's become comparable for continuous star formation at ages of 10 to 20 Myr.

\section{Conclusions}\label{conc}
We present a grid of theoretical stellar \lya\ equivalent widths for evolving stellar populations. While these models have value in their own right, our main motivation for presenting the grid is the need to correct nebular \lya\ emission in star-forming galaxies for underlying stellar absorption. This study builds on the pioneering work of \citet{val93} and \citet{cha93} who had to rely on earlier, less complete model atmospheres, and it extends the exploratory work of \citet{sch08}, who used the latest model atmospheres for a restricted parameter range.

We compiled a new library of non-LTE \lya\ EWÕs based on fully line-blanketed models. The library includes O and B stars with effective temperatures ranging from 15,000 K to 48,500 K, and \logg\ from 1.75 to 4.5. The measured \lya\ EW's range from 5.9 to -32.5 \AA. In particular, we obtained: \lya\ EW between 0.9 and -32.5 \AA\ for effective temperatures of 15,000 K to 29,000 K, and values mostly close to 0 \AA\ for higher effective temperatures due to the canceling effects of the emission and absorption of the P Cygni profiles of these stars. Metallicity has very little effect on the \lya\ EW's.

We implemented the library in Starburst99 and studied the behavior of the \lya\ EW for a range of stellar population properties. We performed a parameter study of instantaneous and continuous star formation at solar metallicity, using the high-mass loss Geneva evolution models \citep{mey94}. We found: (i) The IMF has only minor influence on the {\em stellar} EW when varied within astrophysically plausible limits for both an instantaneous and a constant SFR. (ii) The IMF has also a minor influence on the nebular component of the \lya\ EW for an instantaneous SFR, however, for a continuous SFR, the difference between the two extreme IMF exponents (0.5 and 2.6) in the nebular component of the \lya\ EW is about 100 \AA\ at 3 Myr and about 180 \AA\ at 15 Myr. (iii) When O stars dominate the spectrum (as indicated by the presence of nebular emission lines), stellar \lya\ absorption is always small compared to gaseous \lya\ \emph{(if nebular \lya\ has an escape fraction close to 100\%)}. (iv) Depending on the escape fraction of nebular \lya\ photons, the stellar contribution to the total ranges from negligible to dominant. If the nebular escape fraction is 10\%, the stellar absorption and nebular emission equivalent widths become comparable for continuous star formation at ages of 10 to 20 Myr.

In stark contrast to the often used a stellar absorption correction of 50 to 240 \AA, our models predict: (i) stellar \lya\ EW values of 9 to 18 \AA\ in absorption for an instantaneous burst between the ages of 5 to 15 Myr, and (ii) values of 2 to 8 \AA\ also in absorption for constant star formation in the same age range. Our models provide a realistic description of the stellar \lya\ and should be appropriate to correct the nebular emission for underlying stellar absorption in star-forming galaxies.

\vspace{8mm}
\emph{AKNOWLEDGEMENTS}.
We are grateful to an anonymous referee for a careful reading of the manuscript and many useful suggestions. We are also grateful to John Hillier for kindly providing us with plane-parallel \cgen\ models to check consistency between the results of \cgen\ and \tlus\ modeled atmospheres. Support for this work has been provided by NASA through grant number N-1317 from the Space Telescope Science Institute, which is operated by AURA, Inc., under NASA contract NAS5-26555.

\clearpage

\begin{center}
\begin{deluxetable}{lccccccccc}
\tabletypesize{\scriptsize}
\tablecaption{
Parameters of the models used for the new library. \label{tmods}}
\tablewidth{0pt}
\tablehead{
\colhead{\teff\ [K]} & \colhead{\logg} & \colhead{Code}
}
\startdata
15,000	&1.75, 2.00, 2.25, 2.50, 2.75, 3.00, 3.25, 3.50, 3.75, 4.00, 4.25, 4.50 & \tlus\\
16,000	&2.00, 2.25, 2.50, 2.75, 3.00, 3.25, 3.50, 3.75, 4.00, 4.25, 4.50 & \tlus\\
17,000	&2.00, 2.25, 2.50, 2.75, 3.00, 3.25, 3.50, 3.75, 4.00, 4.25, 4.50 & \tlus\\
18,000	&2.00, 2.25, 2.50, 3.00, 3.25, 3.50, 3.75, 4.00, 4.25, 4.50	& \tlus\\
19,000	&2.25, 2.50, 2.75, 3.00, 3.25, 3.50, 3.75, 4.00, 4.25, 4.50	& \tlus\\
20,000	&2.25, 2.50, 2.75, 3.00, 3.25, 3.50, 3.75, 4.00, 4.25, 4.50 & \tlus\\
21,000	&2.50, 2.75, 3.00, 3.25, 3.50, 3.75, 4.00, 4.25, 4.50 & \tlus\\
22,000	&2.50, 2.75, 3.00, 3.25, 3.50, 3.75, 4.00, 4.25, 4.50 & \tlus\\
23,000	&2.50, 2.75, 3.00, 3.25, 3.50, 3.75, 4.00, 4.25, 4.50 & \tlus\\
24,000	&2.50, 2.75, 3.00, 3.25, 3.50, 3.75, 4.00, 4.25, 4.50 & \tlus\\
25,000	&2.50, 2.75, 3.00, 3.25, 3.50, 3.75, 4.00, 4.25, 4.50 & \tlus\\
26,000	&2.75, 3.00, 3.25, 3.50, 3.75, 4.00, 4.25, 4.50 & \tlus\\
27,000	&2.75, 3.00, 3.25, 3.50, 3.75, 4.00, 4.25, 4.50 & \tlus\\
27,500	&3.00, 3.75 & \cgen\\
27,730	&3.35	& \cgen\\
28,000	&2.75, 3.00, 3.25, 3.50, 3.75, 4.00, 4.25, 4.50 & \tlus\\
29,000	&3.00, 3.25, 3.50, 3.75, 4.00, 4.25, 4.50 & \tlus\\
30,000	&3.00, 3.25, 3.50, 3.75, 4.00, 4.25, 4.50 & \tlus\\
30,000	&3.00, 3.13,  3.50, 4.00, 4.10, 4.20	& \cgen\\
30,270	&3.29	& \cgen\\
30,410	&3.73	& \cgen\\
31,480	&4.06	& \cgen\\
32,210	&3.26	& \cgen\\
32,500	&3.25, 3.50, 3.60, 4.10, 4.25	& \cgen\\
32,660	&3.71	& \cgen\\
33,340	&4.01	& \cgen\\
34,435	&3.52	& \cgen\\
35,000	&3.25, 3.50, 4.10, 4.25 & \cgen\\
36,310	&4.02	& \cgen\\
37,411	&3.38	& \cgen\\
37,500	&3.50, 3.90	& \cgen\\
37,670	&3.77	& \cgen\\
37,760	&3.76	& \cgen\\
39,540	&3.68	& \cgen\\
39,628	&3.93	& \cgen\\
39,994	&3.63	& \cgen\\
40,000	&3.50, 3.90	& \cgen\\
41,020	&4.04	& \cgen\\
41,591	&3.80	& \cgen\\
41,178	&4.02	& \cgen\\
42,560	&3.71	& \cgen\\
42,560	&4.16	& \cgen\\
43,954	&3.98	& \cgen\\
46,130	&4.05	& \cgen\\
48,530	&4.01	& \cgen\\
\enddata
\end{deluxetable}
\end{center}
\clearpage

\begin{center}
\begin{deluxetable}{lccccccccccccc}
\tabletypesize{\small}
\tablecaption{
\lya\ Equivalent Widths\tablenotemark{*} as a function of Effective Temperature and \logg.\label{tewtg}}
\tablewidth{0pt}
\tablehead{
\colhead{\teff [K]} && 
\multicolumn{12}{c}{\lya\ EW [\AA] for \logg\ of:}\\
&&	1.75	&	2.00&	2.25&	2.50&	2.75&	3.00&	3.25&	3.50&	3.75&	4.00&	4.25&	4.50
}
\startdata
15,000	&&	-15.3& -20.6& -18.3& -22.6& -25.5& -25.6& -24.6& -27.9& -29.3& -29.3& -32.6& -32.5\\
16,000	&&	-12.2& -13.8& -16.8& -17.8& -19.8& -20.8& -23.7& -24.2& -25.0& -27.5& 28.8& -29.7\\ 
17,000	&&	-9.6& -11.1& -18.4& -16.0& -16.6& -19.7& -21.7& -21.7& -23.8& -24.6& -26.9& -25.9\\ 
18,000	&&	-3.8& -5.7& -11.4& -13.6& -15.5& -15.4& -17.1& -19.4& -20.3& -24.4& -22.2& -24.3\\ 
19,000	&&	-5.2& -7.0& -8.7& -9.4& -15.7& -14.3& -18.5& -18.6& -20.6& -19.8& -20.7& -24.1\\ 
20,000	&& -5.8& -7.2& -8.7& -11.5& -13.6& -13.6& -16.1& -15.5& -18.01& -17.9& -20.8& -21.7\\
21,000	&& -8.1& -9.2& -10.4& -11.6& -10.8& -12.5& -14.5& -14.4& -16.9& -17.9& -18.8& -20.9\\
22,000	&& -5.0& -6.3& -7.5& -8.7& -10.6& -12.2& -12.4& -14.9& -15.2& -16.9& -18.8& -18.6\\
23,000	&& -1.3& -2.7& -4.0& -5.4& -10.9& -11.0& -12.3& -12.2& -14.8& -14.3& -19.2& -16.2\\
24,000	&& 0.9& -0.9& -2.7& -4.5& -9.0& -10.0& -12.4& -13.5& -15.7& -17.0& -15.8& -18.8\\
25,000	&& -0.7& -2.2& -3.7& -5.3& -6.8& -11.2& -11.4& -11.9& -12.7& -16.3& -16.0& -17.5\\
26,000	&& -4.8& -5.8& -6.8& -7.8& -8.8& -9.6& -10.6& -10.7& -14.1& -12.9& -14.5& 15.9\\
27,000	&& -3.5& -4.4& -5.3& -6.2& -7.0& -10.4& -8.5& -10.0& -12.2& -11.8& -14.0& -13.2\\
28,000	&& -1.0& -2.0& -3.1& -4.2& -5.2& -7.5& -6.6& -8.1& -11.2& -12.8& -11.1& -12.7\\
29,000	&& -3.5& -4.2& -4.9& -5.6& -6.3& -7.0& -6.0& -8.0& -11.0& -12.0& -11.0& -11.2\\
30,000	&& 0.4& -0.1& -0.5& -0.9& -1.3& -1.7& -2.0& -2.6& -3.0& -3.4& -3.8& -4.3\\
32,500	&& 1.5& 1.0& 0.5& -0.0& -0.5& -1.1& -1.4& -2.1& -2.4& -3.1& -2.6& -4.1\\
35,000	&& 5.9& 5.0& 4.0& 3.1& 2.1& 1.2& 0.2& -1.3& -1.7& -2.6& -2.8& -4.5\\
37,500	&& 0.5& 0.4& 0.3& 0.1& 0.0& -0.1& -0.2& -0.4& -0.5& -0.6& -0.7& -0.9\\
40,000	&& 2.0& 1.7& 1.4& 1.1& 0.8& 0.5& 0.2& -0.1& -0.4& -0.7& -1.0& -1.3\\
42,500	&& 2.2& 1.8& 1.5& 1.2& 0.9& 0.5& 0.3& -0.1& -0.4& -0.7& -1.0& -1.4\\
45,000	&& 0.7& 0.6& 0.4& 0.2& 0.1& -0.1& -0.2& -0.4& -0.5& -0.7& -0.9& -1.0\\
48,500	&& -0.3& -0.4& -0.4& -0.5& -0.6& -0.6& -0.7& -0.7& -0.8& -0.9& -0.9& -1.0\\
\enddata
\tablenotetext{*} {The sign convention used for this work is positive for emission and negative for absorption.}
\end{deluxetable}
\end{center}
\clearpage

\begin{center}
\begin{deluxetable}{lllccccccccccc}
\tabletypesize{\small}
\tablecaption{
\lya\ Equivalent Widths Measurements from Observations\tnm{*} as a function of Effective Temperature.\label{tewobs}}
\tablewidth{0pt}
\tablehead{
\colhead{Star} & \colhead{Name} & \colhead{Spectral Type} & \colhead{\lya\ EW[\AA]} 
&  \teff\ [K]\\
\colhead{[HD number]}
}
\startdata
\multicolumn{5}{l}{\emph{Taken from \citet{sav70}}\tnm{a}}\\
11415\tnm{b}&$\epsilon$ Cas	&B3 IVp&	-28&		17,500\\
24398		&$\zeta$ Per		&B1 Ib	&	-25&		21,500\\
24760		&$\epsilon$ Per	&B0.5 V&	-14&		28,000\\
29763		&$\tau$ Tau		&B3 V	&	-29&		17,500\\
32630\tnm{b}&$\eta$ Au		&B3 V	&	-25&		17,500\\
34816		&$\lambda$ Lep	&B0.5 IV&	-14&		27,500\\
35411		&$\eta$ Ori		&B0.5 V&	-20&		28,000\\
35439		& 25 Ori		&B1 Vpe&	-19&		26,000\\
35468		&$\gamma$ Ori	&B2 III	&	-15&		19,750\\
35497\tnm{b}&$\beta$ Tau	&B7 III	&	-35&		12,700\\
36485-86	&$\delta$ Ori	&O9.5 II&	-20&		29,625\\
36512		&$\nu$ Ori		&B0 V	&	-13&		29,500\\
36822		&$\phi'$ Ori		&B0 IV	&	-31&		29,000\\
37022		& 41 Ori		&O7 V	&		&		37,000\\
\hspace{5mm}-41&$\theta$ Ori&O9.5 Vp&	-33&		31,500\\
37043		&$\iota$ Ori		&O9 III	&	-17&		32,250\\
37128		&$\epsilon$ Ori	&B0 Ia	&	-24&		27,500\\
37202		&$\zeta$ Tau	&B2 IVp&	-21&		20,375\\
37468		&$\sigma$ Ori	&O9.5 V&	-16&		31,500\\
37742-43	&$\zeta$ Ori		&O9.5 Ib&	-25&		29,000\\
38771		&$\kappa$ Ori	&B0.5 Ia&	-22&		26,000\\
44743		&$\beta$ CMa	&B1 II	&	-10&		22,625\\
52089		&$\epsilon$ CMa	&B2 II	&	-12&		19,125\\
87901\tnm{b}&$\alpha$ Leo	&B7 V	&	-43&		13,300\\
116658		&$\alpha$ Vir	&B1 V	&	-11&		26,000\\
120315\tnm{b}&$\eta$ Uma	&B3 V	&	-22&		17,500\\
122451		&$\beta$ Cen	&B1 II	&	-12&		22,625\\
127381		&$\sigma$ Lup	&B2 V	&	-20&		21,000\\
127972		&$\eta$ Cen		&B1.5 Vne& -15&		24,000\\
128345\tnm{b}&$\rho$ Lup	&B5 V	&	-36&		15,400\\
129056		&$\alpha$ Lup	&B1 V	&	-14&		26,000\\
132200		&$\kappa$ Cen	&B2 V	&	-17&		21,000\\
133242-43\tnm{b}&$\pi$ Lup	&B5 IV	&	-26&		14,925\\
133955\tnm{b}&$\lambda$ Lup&B3 V	&	-22&		17,500\\
136298		&$\delta$ Lup	&B2 IV	&	-14&		20,375\\
136664		&$\phi^2$ Lup	&B5 V	&	-26&		15,400\\
139365\tnm{b}&$\tau$ Lib	&B2.5 V&	-22&		19,000\\
141637		&1 Sco			&B2.5 Vn&	-30&		19,000\\
142983		&48 Lib			&B8 Ia/Iab&	-12&		11,400\\
143018		&$\pi$ Sco		&B1 V	&	-18&		26,000\\
143275		&$\delta$ Sco	&B0 V	&	-24&		29,500\\
144470		&$\omega^1$ Sco&B1 V	&	-33&		26,000\\
147165		&$\sigma$ Sco	&B1 III	&	-31&		23,750\\
149757		&$\zeta$ Oph	&O9.5 V&	-26&		31,500\\
151890		&$\mu'$ Sco		&B1.5 V&	-20&		24,000\\
158926		&$\lambda'$ Sco	&B1 V	&	-12&		26,000\\
160578		&$\kappa$ Sco	&B2 IV	&	-16&		20,375\\
189103		&$\Theta'$ Sgr	&B3 IV	&	-29&		17,000\\
\vspace{3mm}
209952\tnm{b}&$\alpha$ Gru	&B5 V	&	-33&		15,400\\

\multicolumn{5}{l}{\emph{Taken from \citet{sav74}\tnm{c}}}\\
358 			& $\alpha$ And	&B9 II	&	-45&	10,850\\
10144		& $\alpha$ Eri	&B3 Vp	&	-21&	17,500\\
11415\tnm{b}& $\epsilon$ Cas&B3 Vp&	-25&	17,500\\
19356		& $\beta$ Per	&B8 V	&	-65&	12,300\\
22928		& $\delta$ Per	&B5 III	&	-31&	14,450\\
32630\tnm{b}& $\eta$ Aur	&B3 V	&	-22&	17,500\\
34085		& $\beta$ Ori	&B8 Ia	&	$>$-30	&11,400\\
35497\tnm{b}& $\beta$ Tau	&B7 III	&	-42&	12,700\\
37795		& $\alpha$ Col	&B7 IV	&	-45&	13,000\\
42560		& $\xi$ Ori		&B3 IV	&	-21&	17,000\\
44402		& $\zeta$ CMa	&B2.5 IV&	-20&	18,375\\
6575		& $\chi$ Car		&B3 IVp&	-19&	17,000\\
87901\tnm{b}& $\alpha$ Leo	&B7 V	&	-55&	13,300\\
120315\tnm{b}& $\eta$ UMa	&B3 V	&	-25&	17,500\\
121263		& $\zeta$ Cen	&B2.5 IV&	-10&	18,375\\
125238		& $\iota$ Lup	&B2.5 IV&	-18&	18,375\\
125823		& $\alpha$ Cen	&B7 IIIp&	-23&	12,700\\
128345\tnm{b}& $\rho$ Lup	&B5 V	&	-45&	15,400\\
129116		& 				&B3 V	&	-20&	17,500\\
133242\tnm{b}& $\pi$ Lup	&B5 V	&	-30&	15,400\\
\hspace{7mm}-43\tnm{c}& 	&B5 IV	&		&	14,975\\
133955\tnm{b}& $\lambda$ Lup&B3 V&	-21&	17,500\\
139365\tnm{b}& $\tau$ Lib	&B2.5 V&	-19&	19,000\\
143118		& $\eta$ Lup		&B2.5 IV&	-11&	18,375\\
147394		& $\tau$ Her		&B5 IV&	-35&	14,925\\
155763		& $\zeta$ Dra	&B6 III&	-44&	13,600\\
160762		& $\iota$ Her	&B3 IV&	-21&	17,000\\
175191		& $\sigma$ Sgr	&B3 IV&	-18&	17,000\\
193924		& $\alpha$ Pav	&B2.5 V&	-17&	19,000\\
209952\tnm{b}& $\alpha$ Gru&B7 IV&	-45&	13,000\\
\enddata
\tablenotetext{*} {The signs of the measured EW's were changed from original papers in order to maintain the convention of this work: absorption represented by a negative EW.}
\tablenotetext{a}{The EW's in this table include the blending with neighboring lines within 15\AA\ (the OAO resolution), and also the blend with the interstellar \lya\ line.}
\tablenotetext{b}{These stars are present in both papers, \citet{sav70} and \citet{sav74}.}
\tablenotetext{c}{The \citet{sav74} EW's in this table were those corrected for interstellar absorption and for the blend with neighboring lines.}
\end{deluxetable}
\end{center}
\clearpage

\begin{center}
\begin{deluxetable}{lccccccccc}
\tabletypesize{\scriptsize}
\rotate
\tablecaption{
Comparison of \lya\ Equivalent Widths\tablenotemark{*}. These values where determined for a Salpeter IMF, a stellar population with ages 5$\times10^6$ and 15$\times10^6$ years, and solar metallicity. \label{teqws}}
\tablewidth{0pt}
\tablehead{
\colhead{Authors} && 
\multicolumn{2}{c}{Stellar component EW [\AA]} && 
\multicolumn{2}{c}{Nebular component EW [\AA]} && 
\multicolumn{2}{c}{Total EW [\AA] = stellar + nebular} \\
\cline{3-4}
\cline{6-7}
\cline{9-10}
&& \colhead{5$\times10^6$ years} & \colhead{15$\times10^6$ years}
&& \colhead{5$\times10^6$ years} & \colhead{15$\times10^6$ years}
&& \colhead{5$\times10^6$ years} & \colhead{15$\times10^6$ years}
}
\startdata
\emph{Instantaneous SFR}&&&&&&&&&\\
This work	&&		 -8&	-17&&	     40&		1&&   32& -16\\
\citet{cha93}	&& \nodata&\nodata&& \nodata&\nodata&& 210&  53\\
\citet{val93}	&&-5& 	-15&& 205& 5&& 200& -10\\
\vspace{3mm}
\citet{sch08}	&&		-4&		-20&& 	    47&		3&&   43& -17\\ 

\emph{Constant SFR}&&&&&&&&&\\
This work		&&	-4& -7&& 145& 100&&	141& 93\\
\citet{cha93}	&& \nodata&\nodata&& \nodata&\nodata&& 217& 110\\
\citet{val93}	&& 	-3& -5&& 253&195&& 	250& 190\\ 
\citet{sch08}	&&	-1& -4&& 131& 80&&	130& 76\\
\enddata
\tablenotetext{*} {The sign convention used for this work is positive for emission and negative for absorption.}
\end{deluxetable}
\end{center}
\clearpage

\clearpage

\clearpage

\clearpage

\clearpage

\clearpage

\clearpage

\clearpage

\clearpage

\end{document}